\documentclass{INTERSPEECH2023}

\interspeechcameraready 

\usepackage{arydshln}

\newcommand{\convnext}{{ConvNeXt }}
\newcommand{\convnextt}{{ConvNeXt}}

\title{Adapting a \convnext model to audio classification on AudioSet}
\name{Thomas Pellegrini$^{1,2,3}$, Ismail Khalfaoui-Hassani$^{1,3}$, Etienne Labbé$^{2,3}$, Timothée Masquelier$^4$}
\address{
  $^1$Artificial and Natural Intelligence Toulouse Institute (ANITI), France\\
  $^2$Institut de Recherche en Informatique de Toulouse (IRIT), France\\
  $^3$Université de Toulouse, CNRS, Toulouse INP, UT3, France\\
  $^4$CerCo UMR 5549, CNRS, Université de Toulouse, France}
\email{thomas.pellegrini@irit.fr,ismail.khalfaoui-hassani@univ-tlse3.fr,\\etienne.labbe@irit.fr,timothee.masquelier@cnrs.fr}

\begin{document}

\maketitle
 
\begin{abstract}
In computer vision, convolutional neural networks (CNN) such as ConvNeXt, have been able to surpass state-of-the-art transformers, partly thanks to depthwise separable convolutions (DSC). DSC, as an approximation of the regular convolution, has made CNNs more efficient in time and memory complexity without deteriorating their accuracy, and sometimes even improving it. In this paper, we first implement DSC into the Pretrained Audio Neural Networks (PANN) family for audio classification on AudioSet, to show its benefits in terms of accuracy/model size trade-off. Second, we adapt the now famous ConvNeXt model to the same task. It rapidly overfits, so we report on techniques that improve the learning process. Our best ConvNeXt model reached 0.471 mean-average precision on AudioSet, which is better than or equivalent to recent large audio transformers, while using three times less parameters. We also achieved positive results in audio captioning and audio retrieval with this model\footnote{Our PyTorch source code and checkpoint models are available at \url{https://github.com/topel/audioset-convnext-inf}}.
\end{abstract}
\noindent\textbf{Index Terms}: audio classification, audio tagging, convolutional neural networks, AudioSet, ConvNeXt

\section{Introduction}

In the field of audio classification, the so-called ``PANNs'' for Pretrained Audio Neural Networks, proposed by Kong and colleagues~\cite{Kong2020}, are well-known in the community. Trained on AudioSet~\cite{audioset}, they are an off-the-shelf solution to perform sound event detection (SED) with the 527 event categories present in AudioSet. They are very much used also as feature extractors in all kinds of audio downstream tasks, such as SED focused on a specific set of events, automatic audio captioning, language-based audio retrieval, to name a few.

PANNs are vanilla convolutional neural networks (CNNs).  CNNs are also still used in vision and interest in them has been rising a lot recently, since modernized CNNs~\cite{Liu_2022_CVPR,Rao2022a,Yu2022a} have been shown to outperform vision transformers (\textit{e.g.}, Swin~\cite{Liu_2021_ICCV}) on ImageNet classification and downstream tasks (semantic segmentation and object detection). These modernized CNNs share multiple features that were not used in PANNs: (a) a stem to substantially reduce the input resolution (typically by a factor 4); (b) depthwise separable convolutions (DSC), which drastically reduce the number of parameters and FLOPS, and therefore allow exploring deeper and wider networks, as well as larger kernel sizes ($>3\times3$); (c) residual skip connections; (d) inverted bottlenecks; (e) downsampling layers that use strided convolutions instead of pooling. We suspected that these modern tricks could also be useful for the field of audio classification. We thus decided to adapt one of the most popular modern vision CNNs, named ConvNeXt~\cite{Liu_2022_CVPR}, to this task.



\section{The ConvNeXt architecture/models}

The \convnext models \cite{Liu_2022_CVPR} were proposed as a fully convolutional alternative to the recent attention-based transformer models in computer vision. 
\convnext is based on the use of depthwise separable convolutions (DSC) \cite{Chollet_2017_CVPR}, followed by what is known as an inverted bottleneck \cite{sandler2018mobilenetv2}, responsible for the channel mixing operation. Strictly speaking, a depthwise convolution refers to a convolution where each input channel is convolved by its specific kernel, separate from the others. In a broader definition, they refer to ``grouped'' convolution, where the number of output channels $C_{\text{out}}$ (equal to the number of filters) is a multiple of the number of input channels $C_{\text{in}}$. In this situation, if $C_{\text{out}} = K \times C_{\text{in}}$, the integer $K>1$ is sometimes called the depthwise multiplier.

The ConvNeXt model can be seen as a modernization of the well-known ResNet model \cite{he2016deep} aiming to obtain the best compromise between accuracy on the one hand, flops and throughput on the other hand. To achieve this, the ConvNeXt model has introduced several architectural modifications to the previous ResNet architecture at the macro and micro levels. We have retained most of them for our audio classification task, these are listed below: 
\begin{itemize}
    \item A stem, consisting of a strided convolution where the strides are equal to the kernel sizes (patches), is used as the first convolution layer of the model.  
    \item Four stages with ratios 1:1:3:1 regarding the number of blocks within each stage.
    \item A stage is a sequence of blocks with a depthwise $7\times 7$ convolution followed by an inverted bottleneck, the number of channels and the spatial dimensions are constant within a stage.
    \item An inverted bottleneck contains a pointwise convolution layer that increases the number of channels by four, followed by a GELU activation function~\cite{https://doi.org/10.48550/arxiv.1606.08415}, and another pointwise convolution layer that decreases the number of channels back to its value at the bottleneck input. A residual connection sums the resulting feature maps to the block input.
    \item Separate downsampling layers (regular convolution layers with $2 \times 2$ stride and kernels) between the stages.
\end{itemize}

We tested several ConvNeXt architectures, and our best results were obtained with the Tiny variant where the numbers of blocks within the four stages are 3-3-9-3, and the numbers of channels are 96-192-384-768. Its adaptation to audio classification is described in details in Section~\ref{sec:adapting}. 

\section{AudioSet and evaluation metrics}
AudioSet is a large-scale audio classification dataset, comprised of about 2 million 10-second clips downloaded from YouTube~\cite{audioset}. A set of 527 audio event categories were used to annotate the data. Since several events may co-occur in the clips, multiple labels can be active for a single clip. AudioSet tagging is, thus, a multi-label classification task.
AudioSet is divided into three subsets: the class-wise unbalanced and balanced subsets with respectively 2,042,985 and 22,176 clips, and an eval set of 20,383 clips. We downloaded the data in 2018, and a number of YouTube links were already broken. Our AudioSet data contains 1,921,982 (unbalanced train set), 21,022 (balanced train set) and 19,393 (eval) clips. 

We report the usual evaluation metrics for AudioSet tagging: mean average precision (mAP), mean area under the curve (AUC) and d-prime~\cite{audioset}. They are computed for each class and then averaged (macro-averaging), the higher the better. 

\section{First experiments towards \convnextt}

The use of DSC in audio tagging is not new. In~\cite{drossos2020sound}, for instance, the replacement of regular convolutions by DSCs resulted in large reductions in model complexity, together with performance gains. Nevertheless, we are not aware of this line of work when applied to AudioSet, the largest audio tagging dataset available. In this section, we report two preliminary experiments using slightly modified architectures of PANN's CNN14 and CNN6~\cite{Kong2020}. 
The results, shown in Table~\ref{tab:firstresults}, motivated us to adapt full \convnext vision models to the task.

In~\cite{Kong2020}, CNN14 and CNN6 were trained on a single V100 GPU, with a 32 batch size and for three days. We used their training scripts to train these models and our depthwise variants on eight V100 GPUs in parallel, for 20 hours, with a batch size of 64 samples per GPU, thus, an effective batch size of 512 samples. This larger batch size is responsible for significant gains in performance, when we compare the CNN6 and CNN14 results reported in~\cite{Kong2020}, and ours, referred to as ``CNN14 (rep.)'' and ``CNN6 (rep.)'' hereafter.

\begin{table}[htbp]
\caption{First results on the AudioSet test set, reproducing CNN14 and CNN6, and testing variants using depthwise convolutions. }
\begin{center}
\begin{tabular}{lcccc}
\toprule
Model &     \# params &     mAP &   AUC &   d-prime \\
\midrule
CNN14~\cite{Kong2020} & 80.8M & .431 & .973 & 2.732 \\
CNN14 (rep.) & 80.8M & .441 & .972 & 2.971 \\ 
CNN14Sep & 30.5M & .436 & .973 & 2.984 \\
\midrule
CNN6~\cite{Kong2020} &  \hphantom{1}4.8M & .343 & .965 & 2.568 \\
CNN6 (rep.) &   \hphantom{1}4.8M & .350 & .967 & 2.802 \\
CNN6Next  &     \hphantom{1}3.3M & .427 & .972 & 2.957 \\
\bottomrule
\end{tabular}
\label{tab:firstresults}
\end{center}
\end{table}
\vspace*{-4mm}
\subsection{CNN14Sep: CNN14 with depthwise convolutions}

CNN14 consists of six blocks with regular convolution layers with kernels of size $3 \times 3$, followed by global pooling and two fully-connected layers. Each convolution block has two convolution layers. The first one doubles the number of input channels, in order to compensate for the $2\times 2$ spatial pooling operation preceding the block. The second convolution layer keeps the number of channels unchanged. We replaced this second layer by a depthwise convolution one. We refer to this CNN14 variant as CNN14Sep. The number of learnable parameters dropped from 80.8 to 30.5 million. CNN14Sep achieved a 0.436 mAP, which is better than the 0.431 value reported in~\cite{Kong2020}, but 0.005 lower than our reproduced CNN14, trained with a large batch size.

\subsection{CNN6Next: CNN6 with ConvNeXt-like blocks}
\vspace*{-4mm}
\begin{table}[th]
\caption{Architecture of PANN's CNN6 and our variant CNN6Next. DW: depthwise, K: depthwise multiplier, BN and LN: Batch and Layer Normalization. }
\begin{center}
\begin{tabular}{|c|c|}
\hline
CNN6 & CNN6Next \\
\hline
\multicolumn{2}{|c|}{Log-Mel spectrograms} \\
\multicolumn{2}{|c|}{1000 frames $\times$ 64 mel bins} \\
\multicolumn{2}{|c|}{BN on the 64 mel features} \\
\hline
$5\times 5$@64  & $7\times 7$@64  \\
BN, ReLU   & LN, $1\times 1$@256 \\
        & GeLU, $1\times 1$@64 \\
\hline
\multicolumn{2}{|c|}{Avg Pooling $2 \times 2$} \\
\hline
$5\times 5$@128 & $7\times 7$@128 DW-K=2 \\
BN, ReLU & LN, $1\times 1$@512 \\
        & GeLU, $1\times 1$@128 \\
\hline
\multicolumn{2}{|c|}{Avg Pooling $2 \times 2$} \\
\hline
$5\times 5$ @ 256 &  $7\times 7$@256 DW-K=2 \\
BN, ReLU & LN, $1\times 1$@1024 \\
        & GeLU, $1\times 1$@256  \\
\hline
\multicolumn{2}{|c|}{Avg Pooling $2 \times 2$} \\
\hline
$5\times 5$ @ 512 & $7\times 7$@512 DW-K=2 \\
BN, ReLU & LN, $1\times 1$@2048 \\
        & GeLU, $1\times 1$@512 \\
\hline
\multicolumn{2}{|c|}{Global Pooling}\\
\hline
\multicolumn{2}{|c|}{FC 512, ReLU}\\
\multicolumn{2}{|c|}{FC 527, sigmoid}\\
\hline
\end{tabular}
\label{tab:cnn6next}
\end{center}
\end{table}
\vspace*{-6mm}

We ran experiments with the smaller PANN architecture CNN6, in which we replaced the regular convolution blocks with simplified \convnext  convolution blocks, as shown in Table~\ref{tab:cnn6next}. We needed to slightly adapt these blocks, in order to cope with the number of channels that is doubled at each convolution/block layer in CNN6. Thus, instead of having a fully depthwise $7 \times 7$ convolution layer, we used a depthwise layer with a depthwise multiplier of two ($K=2$). We kept the inverted bottlenecks and removed the residual connections (not possible to keep them since the number of channels is doubled at each block). We also simplified the blocks by using $2\times 2$ average pooling instead of the \convnext downsampling convolution layers. A 0.427 mAP value was obtained with CNN6Next, although it has 3.3M parameters only. This shows that a large gain can be obtained by modernizing the vanilla CNN6. Nevertheless, CNN6Next is not scalable, for the two following reasons: residual connections are not possible in this architecture and adding more blocks doubles the number of channels at each new one, leading to strong overfitting.

\section{Adapting a ConvNeXt architecture to audio classification\label{sec:adapting}}
We adapted ConvNeXt to audio classification on AudioSet by changing the stem and the head of the model. This allows us to eventually take advantage of the available checkpoints pretrained on ImageNet, as bootstrap initializations. 

The so-called stem layer, a term borrowed from transformers, is the first convolution layer in the \convnext models. It is responsible for reducing the spatial dimensions of the input, by outputting ``patches'', \textit{i.e.} feature maps obtained from disjoint subparts of the input images. In the original vision \convnextt s, the size of these output patches was set to $56\times 56$, in order to be consistent with the $7\times 7$ convolution layers that follow, and also with the three downsampling layers that reduce the spatial dimensions by a factor 8 in total. In our task, the input spectrograms are not at all squared, they are much lengthier in time than in frequency, $1000 \times 64$ in the case of the PANN's CNNs. With \convnextt, our best results were obtained with $1000 \times 224$-sized spectrograms, reduced to $252\times 56$ feature maps with our audio-adapted stem. We tested three other resolutions: the original squared $56\times 56$, $122\times 122$ and $504\times 56$, they all obtained worse results than $252\times 56$. For the classification head, we replaced the 1000-dimensional fully-connected layer by a 527-d one, adapted to the 527 AudioSet target categories. 

\section{Experiments on AudioSet}


\subsection{Audio augmentation methods}

We used three audio augmentations when training \convnext models: SpecAugment~\cite{park2019specaugment} and mixup~\cite{zhang2018mixup}, both as implemented in PANN, and we added \textit{Speed Perturbation}~\cite{ko2015audio,cances2022comparison}. SpecAugment is used to randomly drop one or several time or frequency stripes from the spectrograms, meaning that the values in the selected stripes are artificially set to zero. In PANN, the spectrograms have 1000 time frames and 64 log mel frequency bins. During training, the number of dropped stripes is randomly drawn between zero and two for each minibatch. The maximum stripe width is 64 frames in time, and 8 bins in frequency. We kept this setting with \convnextt, except that the maximum stripe width was increased to 28 bins in frequency, since we chose \convnextt's spectrograms to have 224 frequency bins reduced to 56 by the stem. Mixup in PANN operates a convex linear summation of two samples and their labels, with a weight drawn for each example in a minibatch, using a beta probability distribution of parameters $\alpha=\beta=1$. Finally, speed Perturbation refers to resampling the raw audio signals either up (nearest-neighbor upsampling) or down (decimation) according to a rate chosen randomly within $[0.5, 1.5]$. The resulting waveform is, thus, shorter or longer than the original one. Padding or cropping is randomly applied at the start and the end of the stretched signal in order to keep the signal duration constant. 


\subsection{Training setup}

We trained the models with an effective batch size of 512 audio samples, on eight V100-32GB GPUs, for 75k iterations (corresponding to about 20 hours of training). We used the AdamW optimizer and a ``one-cycle'' learning rate (LR) scheduler with a maximum LR value of 4e-3, reached after 30\% of the total number of training steps (22.5k iterations). Weight decay (WD) was crucial to succeed in training \convnext-Tiny models, which strongly overfit otherwise. Several values were tested and a 0.05 WD value was chosen. A 0.4 drop path~\cite{huang2016deep} level was used, also very useful to reduce overfitting.

\subsection{Results on the AudioSet test set}

Table~\ref{tab:results} shows the number of parameters, FLOPs and throughput (samples / s) when available, and the mAP values of our best ConvNeXt model (last line), and of models from the literature: PANN's CNN14, two audio transformers: Audio Spectrogram Transformer (AST,~\cite{gong2021ast}), PaSST-S~\cite{koutini2021efficient}, and a ConvNext-Tiny, implemented in the JAX language in an audio processing framework called Audax\footnote{https://github.com/SarthakYadav/audax}. We report results from the literature obtained with single models, without ensembling methods. Our model achieved a 0.471 mAP (0.973 AUC, 3.071 d-prime), outperforming AST and CNN14 (the original and the reproduced one), and is on par with PaSST, although having about three times less parameters and a throughput about twice as fast. Compared to CNN14, the decrease of throughput observed in \convnext is due to the larger size of the input spectrograms processed by the stem (224 bins instead of 64). Finally, the ConvNeXt-Tiny from Audax was reported to only reach 0.402 mAP, a much lower value than ours. This large gap is probably due to a lack of adaptation to the task, in particular they kept the squared $56 \times 56$ stem output resolution.

We would like to mention that Schmid and colleagues~\cite{schmid2022efficient} achieved better results, in particular a 0.483 mAP with a large MobileNetV3 model (MobileNet40\_as\_ext, 68.4 million parameters). Nevertheless, their models were trained in a completely different way than ours: knowledge distillation used to mimic the predictions of a large ensemble of PaSST models. We would have to use the same learning framework with our models for the sake of a fair comparison. 

Finally, we tested the larger model \convnextt-Small (49.9M parameters) variant. It differs from \convnextt-Tiny only by the number of blocks of the third stage: 3-3-27-3, instead of 3-3-9-3. Although trained with larger values for WD (0.1) and drop-path (0.8), \convnextt-Small still suffers from overfitting and achieved 0.458 mAP only. 

\begin{table}[htbp]
\caption{Mean average precision (mAP) on AudioSet. The throughput was calculated at
inference, on a batch consisting of 64 samples of size 320 000 (10 s), using a single V100-32GB gpu. }
\vspace*{-4mm}
\begin{center}
\begin{tabular}{llccc}
\toprule
Model & \# params & FLOPs & Throughput &   mAP \\
        & &  & (samples/s) & \\
\midrule
CNN14~\cite{Kong2020} & 80.7M & \textbf{21.2G} & \textbf{378.2} & .431 \\
AST~\cite{gong2021ast}       & 88.0M & & & .459 \\
PaSST-S~\cite{koutini2021efficient}  & 87.0M & & 88.7 & \textbf{.471}  \\
\\
ConvNeXt-Tiny & 28.2M & & & .402\\
(JAX, audax) \\ 
\hdashline \\
ConvNeXt-Tiny  & \textbf{28.2M} & \textbf{21.1G} & \textbf{153.6} & \textbf{.471}\\
(ours) & & &\\
\bottomrule
\end{tabular}
\label{tab:results}
\end{center}
\end{table}
\vspace*{-8mm}





\subsection{Influence of pretraining on ImageNet\label{sec:imagenet}}

Our best result on AudioSet, reported in Table~\ref{tab:results}, was obtained using a model initialization pretrained on ImageNet1K. A randomly-initialized model achieved 0.462 mAP (0.970 AUC, 3.031 d-prime), which is about 1\% absolute worse than with pretraining. This result is inline with the AST and PaSST works, in which models pretrained on ImageNet were better initializations than random ones. 

\section{Downstream tasks}
In this section, we report very encouraging results in two downstream tasks, obtained with our \convnextt-Tiny model, trained on AudioSet, and used frozen.

\subsection{Automated Audio Captioning (AAC) on AudioCaps}

\begin{table}[tbhp]
    \centering
    \caption{Audio Captioning results on the AudioCaps test set.}
    \label{tab:aac_results}

    \begin{tabular}{lccc}
        \toprule
        Model & mAP & SPIDEr & FENSE \\
              &     & (CIDEr-D/SPICE) & \\ 
        \midrule
        SOTA~\cite{https://doi.org/10.48550/arxiv.2210.17143} & N/A & {.475 (.769/.181)} & {N/A} \\ 
        \midrule
        CNN14~\cite{Kong2020} & .647 & .435 (.697/.174) & .610 \\
        CNN14 (rep.) & .727  & .451 (.725/.177) & .620 \\ 
        \hdashline \\
        ConvNeXt (ours) & \textbf{.749} & \textbf{.471} (\textbf{.760}/\textbf{.182}) & \textbf{.638} \\
        \bottomrule
    \end{tabular}
\end{table}

AAC  aims to build models that can produce a sentence written in natural language, that describes the content of an audio recording. 
Most AAC systems employ encoder models pretrained on AudioSet to better recognize and describe sound events. In this work, we use a decoder architecture similar to~\cite{labbe2022_t6a, xinhao2021_t6}. We extract a sequence of 31 audio embedding for each audio file, using either CNN14 or our ConvNeXt-Tiny model, without their last pooling and classification layers. The encoder weights are frozen during training. We add a projection layer after the encoders to match the input embedding dimension of the decoder part. The decoder is a standard transformer decoder~\cite{https://doi.org/10.48550/arxiv.1706.03762} with 6 layers, 4 attention heads per layer, a global embedding size set to 256, a global dropout probability of 0.2 and GELU. The final model contains 12.4M trainable parameters and 79.7M frozen parameters. 
We used AudioCaps~\cite{kim-etal-2019-audiocaps}, which is the largest audio-language dataset publicly available. The audio files are 10-second clips extracted from YouTube. Our version of the dataset contains 46 230 audio files in the training subset, 464 in the validation and 912 in the test subsets.

During inference, we use the standard beam search algorithm with a beam size set to two. We report several standard metrics used in AAC. CIDEr-D~\cite{vedantam_cider_2015} is the cosine similarity scores between the TF-IDF scores of the n-grams, SPICE~\cite{anderson_spice_2016} computes an F-Score value over the semantic propositions extracted from the sentences and SPIDEr~\cite{liu_improved_2017} is the mean of CIDEr-D and SPICE. FENSE~\cite{zhou_can_2022} combines two scores: the first one being the cosine similarity between SBERT embeddings extracted from the sentences, the second one is the binary detection made by a Fluency Error Detector trained to detect the common errors made by AAC systems. 
We also report the mAP scores of each encoder on the AudioCaps test subset.

The results in Table~\ref{tab:aac_results} show a clear improvement in AAC scores when using a stronger pre-trained encoder. The best encoder is ConvNeXt-Tiny, which achieves a 0.471 SPIDEr score, a value very close to the SOTA method which reach a SPIDEr value of 0.475. Moreover, our model contains only 40.6M parameters (28.2M frozen and 12.4M trainable) compared to the 108M trainable parameters used in the SOTA system.

\subsection{Language-based Audio Retrieval on Clotho v2.1}

This multimodal task is, to some extent, the reverse task of AAC. It aims to retrieve and rank audio recordings based on a queried caption, formulated as a free-form sentence written in natural language.
The Clotho v2.1 dataset~\cite{drossos2020clotho} was used for this task in the Detection and Classification of Acoustic Scenes and Events (DCASE) 2022 challenge. The development set contains 3839 audio clips with 19 195 reference captions for training, and 1045 audio recordings with 5225 captions in the ``Development-testing'' subset, on which we report our results in Table~\ref{tab:lbar}. The main evaluation metric is a standard metric in information retrieval: mean Average Precision at top-10 (mAP@10). We also report recall values at top-10 (R@10). All the systems use two pretrained neural networks: an audio encoder (PANN and PaSST pretrained on AudioSet) and a text encoder (sentence BERT variants, pretrained on sentence similarity tasks). The state-of-the-art system~\cite{mei2022_t6b} uses PANN's CNN14 and BERT. We were not able to reproduce their result, instead we used another system, called PaSST-MPnet~\cite{pellegrini2022language}, in which PaSST was used to encode the audio recordings using the 527-AudioSet class logits as embeddings. MPnet is a sentence BERT model that provides 768-dimensional embeddings for sentences. The audio logits embeddings are projected onto the MPnet ones with a single linear layer. We used the code provided by the authors, and simply replaced the logit embeddings of PaSST by those predicted with our ConvNeXt-tiny trained on AudioSet. As shown in Table~\ref{tab:lbar}, PaSST and ConvNeXt performed similarly, with 0.229 and 0.230 mAP@10, respectively. An enhanced variant named PaSST+tags was proposed in~\cite{pellegrini2022language}, where the MPnet textual embeddings of the AudioSet tags are combined to the audio logit embeddings, in the audio encoding part of the system. When replacing PaSST with ConvNeXt in this second system, we observed a significant improvement in mAP@10, with a 0.240 value for ConvNeXt+tags, compared to 0.234 for PaSST+tags.

\begin{table}[htbp]
    \centering
    \caption{Audio retrieval results on the Clotho Dev-test split.}
    \label{tab:lbar}
    \begin{tabular}{lccc}
    \toprule
    Audio encoder  & \# params & mAP@10 & R@10 \\
    \midrule
    SOTA~\cite{mei2022_t6b} & 195M & .260 & .530 \\ 
    \midrule
    PaSST-MPNet~\cite{pellegrini2022language}        & 196M & .229 & .482 \\ 
    \hspace{2mm} + tags~\cite{pellegrini2022language}    & 196M & .234 & .485 \\
    \hdashline \\
    ConvNeXt-MPNet (ours)       & 146M & .230 & .488 \\ 
    \hspace{2mm} + tags (ours)  & \textbf{146M} & \textbf{.240} & \textbf{.488} \\ 
    \bottomrule
    \end{tabular}
\end{table}
\vspace*{-4mm}

\section{Conclusions}
In this work, we wanted to revisit CNNs applied to audio classification under the new light shed by CNNs recently modernized in computer vision. We show that a ConvNeXt-Tiny model, adapted to audio tagging on AudioSet, achieved the same precision as a recent transformer, with about three times less parameters. This model achieved very good results on two downstream tasks, audio captioning and language-based audio retrieval. We plan to continue this line of work, but using other learning paradigms, such as knowledge distillation and self-supervision to further explore the capabilities of this type of models.

\section{Acknowledgments}

This work was partially supported by the French ANR agency within the LUDAU project (ANR-18-CE23-0005-01) and the French "Investing for the Future PIA3" AI Interdisciplinary Institute ANITI (Grant agreement ANR-19-PI3A-0004). This work was performed using HPC resources from GENCI-IDRIS (Grant 2022-AD011013587). 

\bibliographystyle{IEEEtran}
\bibliography{mybib}

\end{document}